\documentclass[aps,amsmath,preprint,superscriptaddress,nofootinbib,prd]{revtex4-1}
\usepackage{graphicx}
\usepackage{bm}
\usepackage{color}
\usepackage{amsmath,amssymb}	
\usepackage{hyperref}
\usepackage{setspace}

\usepackage[english]{babel}
\usepackage[utf8x]{inputenc}
\usepackage[colorinlistoftodos]{todonotes}
\usepackage{slashed}

\usepackage{physics}
\usepackage{braket}

\def\simgt{\mathrel{\lower2.5pt\vbox{\lineskip=0pt\baselineskip=0pt
           \hbox{$>$}\hbox{$\sim$}}}}
\def\simlt{\mathrel{\lower2.5pt\vbox{\lineskip=0pt\baselineskip=0pt
           \hbox{$<$}\hbox{$\sim$}}}}
\def\simprop{\mathrel{\lower3.0pt\vbox{\lineskip=1.0pt\baselineskip=0pt
             \hbox{$\propto$}\hbox{$\sim$}}}}
\def\tr{\mathop{\rm tr}}

\begin{document}

\title{Chiral Composite Asymmetric Dark Matter}
\author{Masahiro~Ibe}
\email[e-mail: ]{ibe@icrr.u-tokyo.ac.jp}
\affiliation{Kavli IPMU (WPI), UTIAS, University of Tokyo, Kashiwa, Chiba 277-8583, Japan}
\affiliation{ICRR, University of Tokyo, Kashiwa, Chiba 277-8582, Japan}
\author{Shin~Kobayashi}
\email[e-mail: ]{shinkoba@icrr.u-tokyo.ac.jp}
\affiliation{ICRR, University of Tokyo, Kashiwa, Chiba 277-8582, Japan}
\author{Keiichi~Watanabe}
\email[e-mail: ]{keiwata@icrr.u-tokyo.ac.jp}
\affiliation{ICRR, University of Tokyo, Kashiwa, Chiba 277-8582, Japan}

\begin{abstract}
The asymmetric dark matter (ADM) scenario solves the baryon-dark matter coincidence problem when the dark matter (DM) mass is of $\order{1}$\,GeV. 
Composite ADM models based on QCD-like strong dynamics are particularly motivated since the strong dynamics naturally provides the DM mass of $\order{1}$\,GeV and the large annihilation cross-section simultaneously. 
In those models, the sub-GeV dark photon often plays an essential role in transferring the excessive entropy in the dark sector into the visible sector, i.e., the Standard Model sector. 
This paper constructs a chiral composite ADM model where the $U(1)_D$ gauge symmetry is embedded into the chiral flavor symmetry. 
Due to the dynamical breaking of the chiral flavor symmetry, the model naturally provides the masses of the dark photon and the dark pions in the sub-GeV range, both of which play crucial roles for a successful ADM model.
\end{abstract}

\date{\today}
\maketitle
\preprint{IPMU 21-0031}

\renewcommand{\thesection}{\Roman{section}}
\section{Introduction} 

The observed mass density of dark matter (DM) is about five times larger than that of the baryons in the present Universe. 
The coincidence seems to require a conspiracy between a mechanism that determines the DM abundance and baryogenesis since slight baryon asymmetry generated in the early Universe determines the present baryon abundance.
The asymmetric dark matter (ADM) scenario is an attractive scenario motivated by this coincidence problem~\cite{Nussinov:1985xr,Barr:1990ca,Barr:1991qn,Dodelson:1991iv,Kaplan:1991ah,Kuzmin:1996he,Foot:2003jt,Foot:2004pq,Hooper:2004dc,Kitano:2004sv,Gudnason:2006ug,Kaplan:2009ag}~(\cite{Davoudiasl:2012uw,Petraki:2013wwa,Zurek:2013wia} for reviews).
The ADM scenario assumes that the dark sector, which contains the DM, shares matter-antimatter asymmetry with the visible sector, i.e., the Standard Model sector.
With the asymmetry similar to the baryon asymmetry, the ADM scenario naturally explains the coincidence problem if 
the dark matter mass is of $\order{1}$\,GeV.

Composite ADM models based on QCD-like strong dynamics~\cite{Foot:2003jt,Foot:2004pq,Berezhiani:2005ek,Alves:2009nf,An:2009vq,Alves:2010dd,Gu:2012fg,
Buckley:2012ky,Detmold:2014qqa,Gu:2014nga,Fukuda:2014xqa,Garcia:2015toa,Lonsdale:2018xwd,Ritter:2021hgu,Zhang:2021orr} are particularly motivated by two reasons.
First, strong dynamics naturally provides a large annihilation cross-section, which is an important feature for successful ADM so that the symmetric components annihilate away and  the asymmetric component becomes the dominant component of the DM abundance.
Second, it can also provide the DM mass in the GeV range through dimensional transmutation.

The composite ADM models contain numerous particles since the models are based on strong dynamics of a non-Abelian gauge theory.
Thus, if the dark sector decouples from the visible sector,
the lightest particles in the dark sector overclose the Universe or result in too large contribution to the effective number of neutrinos, $N_\mathrm{eff}$~\cite{Blennow:2012de} (see also Ref.~\cite{Ibe:2019gpv}).
Therefore, we need a light portal particle
that transfers the entropy of the dark sector into the visible sector to avoid the problems.%
\footnote{We assume that the two sectors are in thermal equilibrium at the temperature around the baryogenesis.
It is also possible to avoid the problems without a portal particle if the dark sector temperature is lower than that of the visible sector when the lightest dark sector particle is massless.}
One candidate of such a portal particle is a dark photon
which couples to the QED photon through a kinetic mixing~\cite{Holdom:1985ag}.
As demonstrated in Ref.~\cite{Ibe:2018juk},
the dark photon with a mass in the sub-GeV range successfully transfers the entropy of the dark sector into the visible sector.

In this paper, we construct a model that naturally provides the dark photon mass in the sub-GeV range while the mass of the composite ADM is in the GeV range.
For this purpose, we rely on the dynamical generation of the dark photon mass proposed in Refs.~\cite{Harigaya:2016rwr,Co:2016akw}.
There, the $U(1)_D$ gauge symmetry is embedded into the chiral flavor symmetry. Due to the dynamical breaking of the chiral flavor symmetry, the model naturally provides the masses of the dark photon and the dark pions in the sub-GeV range.

The organization of the paper is as follows. 
In Sec.\,\ref{sec:MODEL}, we first review the composite ADM model in Ref.~\cite{Ibe:2018juk} and the mechanism in Refs.~\cite{Harigaya:2016rwr,Co:2016akw} which gives the dark photon mass dynamically.
Next, we apply this dynamical mechanism to a simple composite ADM model.
Finally, we discuss a concrete example where the asymmetries in the dark sector 
and the visible sector are thermally distributed through higher-dimensional operators.
In Sec.\,\ref{sec:phenomenology}, we discuss the phenomenology of the models.
The final section is devoted to our conclusions.

\section{MODEL}
\label{sec:MODEL}

In this section, we first review the composite ADM model in Ref.~\cite{Ibe:2018juk}.
Next, we review the mechanism by which the dark photon obtains the mass dynamically.
After that, we construct a chiral composite ADM model with two flavors of the dark quarks as a simple example.
Finally, we construct a three-flavor model compatible with a $B-L$ portal interactions which distribute a $B-L$ asymmetry between the visible sector and the dark sector. 
Hereafter, $N_f$ represents the number of the flavors of dark quarks.

\renewcommand{\thesubsection}{\Alph{subsection}}
\subsection{A Model of (non-chiral) Composite ADM}
\label{sec:CADM}

In this subsection, we review the composite ADM model in
Ref.~\cite{Ibe:2018juk}.
The model is based on $ SU(3)_{D} \times U(1)_{D} $ gauge symmetry with 
$N_f = 2$.  
The dark quarks are in the fundamental representations of $SU(3)_{D}$ and have $U(1)_{D}$ and $ U(1)_{B-L}$ charges like the up-type and the down-type quarks in the visible sector. 
In Tab.\,\ref{tab:tab two flavor}, we show the charge assignment of the dark quarks, where all the (anti-)quarks are given as the left-handed Weyl fermions.
Hereafter, we put the primes on the fields and the parameters in the dark sector.
As is evident from the table,
$SU(3)_{D} $ and $ U(1)_{D} $ symmetries are analogous to the visible QCD and QED,
and hence, we call them dark QCD and dark QED, respectively.
Incidentally, let us emphasize that the absence of the dark weak interaction and the dark neutrinos is a quite important feature for successful cosmology because the dark neutrinos, if present, can be the lightest dark sector particle and lead to too large $N_{\mathrm{eff}}$.

\begin{table}[t]
 \begin{center}
  \begin{tabular}{|c|c|c|c|} \hline
           & $SU(3)_{D}$ & $U(1)_{D}$ & $U(1)_{B-L}$ \\ \hline
    $U^{'}$ & \bf{3} & 2/3 & 1/3 \\
    $D^{'}$ & \bf{3} & $ -1/3 $ & 1/3 \\
    $\bar{U}^{'}$ & $\bar{\bf{3}}$ & $ -2/3 $ & $ -1/3 $ \\ 
    $\bar{D}^{'}$ & $\bar{\bf{3}}$ & 1/3 & $ -1/3 $ \\ \hline
  \end{tabular}
   \caption{\small\sl The charge assignment of the dark quarks in the vector-like QED model in Ref.\,\cite{Ibe:2018juk}.
   Both the dark quarks and the dark anti-quarks are given as the left-handed two-component Weyl fermions.}
  \label{tab:tab two flavor}
 \end{center} 
\end{table}

Since the dark quarks are in the vector-like representation, they have masses,
\begin{align}
\label{eq:Lmass}
\mathcal{L}_{\mathrm{mass}} = m_{U'} \bar{U}^{\prime} U^{\prime} + m_{D'} \bar{D}^{\prime} D^{\prime} + \mathrm{h.c.}
\end{align}
Here, we assume that the dark quark mass parameters $m_{U',D'}$ are smaller than the dynamical scale of dark QCD, $\Lambda^{\prime}_{D}$. 
Below $\Lambda^{\prime}_{D}$, the dark quarks are confined into the dark mesons and dark baryons as in QCD.
The lightest dark mesons are given by
\begin{align}
\label{eq:darkpions}
\pi^{\prime 0} \propto U^{\prime} \bar{U}^{\prime} - D^{\prime} \bar{D}^{\prime} ~ , ~~ \pi^{\prime +} \propto U^{\prime} \bar{D}^{\prime} ~ , ~~ \pi^{\prime -} \propto D^{\prime} \bar{U}^{\prime} \ ,
\end{align}
which are analogous to the pions in QCD.
The lightest dark baryons are also given by,
\begin{align}
p^{\prime} \propto U^{\prime} U^{\prime} D^{\prime} ~ , ~~ \bar{p}^{\prime} \propto \bar{U}^{\prime} \bar{U}^{\prime} \bar{D}^{\prime} ~ , ~~ 
n^{\prime} \propto U^{\prime} D^{\prime} D^{\prime} ~ , ~~ 
\bar{n}^{\prime} \propto \bar{U}^{\prime} \bar{D}^{\prime} \bar{D}^{\prime} ,
\end{align}
which are analogues of the proton and the neutron, respectively.
By the assumption of $m_{U',D'}\ll \Lambda'_D$, the dark pion are much lighter than the dark baryons as in the case of QCD.

Since the $B-L$ symmetry 
is well approximate symmetry
within the dark sector,
the lifetime of the dark baryons 
can be much longer than the age of the Universe~\cite{Fukuda:2014xqa}.
The dark baryons annihilate 
into the dark pions, and hence, the relic density of 
the symmetric component
is subdominant due to the large annihilation cross-section.
Hence, the present dark baryon density is dominated by the asymmetric component as in the case of the visible baryon.
For a given ratio of the $B-L$ asymmetry in the visible sector to that of the dark sector, $A_\mathrm{DM}/A_{\mathrm{SM}}$, the observed dark matter abundance is explained for
\begin{align}
\label{eq:DMmass}
m_{\mathrm{DM}} \simeq \frac{\Omega_{\mathrm{DM}}}{\Omega_{\mathrm{B}}} \frac{A_{\mathrm{B}}}{A_{\mathrm{SM}}} \frac{A_{\mathrm{SM}}}{A_{\mathrm{DM}}} \times m_{\mathrm{N}} \ ,
\end{align}
with the observed values, $\Omega_{\mathrm{DM}}h^2 =  0.120\pm 0.001$ and $\Omega_B h^2 =  0.0224\pm 0.0001$~\cite{Aghanim:2018eyx}.
Here, $A_\mathrm{B}/A_{\mathrm{SM}}=30/97$ is the ratio between the baryon asymmetry and the $B-L$ asymmetry in visible sector~\cite{Harvey:1990qw},%
\footnote{We assume that the top quark decouples earlier than the sphaleron processes~\cite{Ibe:2011hq}.}
and $m_\mathrm{N} \simeq 0.94\,$GeV is the nucleon mass. 
Thus, for $A_\mathrm{DM}/A_{\mathrm{SM}}= \order{1}$\, the dark matter mass should be of $\order{1}$\,GeV.
Such baryon masses are achieved when the dynamical scale of the dark QCD is of $\order{1}$\,GeV.

For example, in a class of models in which the $B-L$ symmetry in the visible and the dark sectors are thermally distributed through the higher dimensional operator, the ratio is given by $A_{\mathrm{DM}}/A_{\mathrm{SM}}=22N_f/237$~\cite{Fukuda:2014xqa} (see also the Appendix\,\ref{sec:calc of asym}).%
\footnote{Here, we neglected the conservation of the $U(1)_D$ charge.
The correct ratio is $66/395$ for the charge assignment in Tab.\,\ref{tab:tab two flavor}, which is not very different from $22N_f/237$ with $N_f=2$.
}
In this case, the mass of the DM (i.e. the dark baryons) is required to be
\begin{align}
    m_\mathrm{DM} \simeq \frac{17}{N_f}\,\mathrm{GeV}\ ,
\end{align}
where $N_f = 2$ is the minimal choice.

The dark pions are also produced abundantly in the thermal bath, 
which could cause cosmological problems if they remain in the present Universe.
Those problems are solved by the presence of the dark photon lighter than a half of the pion masses.
The dark neutral pion immediately decays into a pair of the dark photons due to the chiral anomaly. 
The stable dark charged pions also annihilate into a pair of the dark photons, and hence, their relic abundance is subdominant. 

The dark photon eventually  
decays into a pair of the electron and the positron  through the kinetic mixing with the visible photon given by,
\begin{align}
\mathcal{L}_{\gamma^{\prime}} =
-\frac{1}{4}F^{\prime\mu\nu}F'_{\mu\nu}+\frac{\epsilon}{2} F_{\mu \nu} F^{\prime \mu \nu} + \frac{1}{2} m^{2}_{\gamma^{\prime}}A^{\prime}_{\mu} A^{\prime \mu}\ .
\end{align}
Here, $m_{\gamma^{\prime}}$ denotes the mass of the dark photon, $A_{\mu}^{\prime}$, $\epsilon$ is the kinetic mixing parameter, and $F_{\mu \nu}$ and $F_{\mu \nu}^{\prime}$ are the field strengths of the visible and the dark photons, respectively.
As discussed in Ref.\,\cite{Ibe:2018juk},
the dark photon successfully transfers the excessive energy/entropy in the dark sector into the visible sector for 
$m_{\gamma'} = \order{10\mbox{--}100}$\,MeV 
and $\epsilon =\order{10^{-8}\mbox{--}10^{-11}}$.%
 \footnote{See Refs.~\cite{Ibe:2018tex,Ibe:2019ena} for a possible origin of the tiny kinetic mixing parameters. }

\renewcommand{\thesubsection}{\Alph{subsection}}
\subsection{Dynamical Generation of Dark Photon Mass}
\label{sec:Dynamical}
In the above discussion, we have not specified the origin of the dark photon mass.
The simplest possibilities are to introduce a Higgs boson or to assume the St\"uckelberg model~\cite{Ruegg:2003ps}.
In these models, however, we require parameter tuning so that the dark photon mass is of $\order{10\mbox{--}100}$\,MeV while the DM mass is of $\order{1}$\,GeV.
To avoid such parameter tuning, we apply a mechanism which generates a dark photon mass due to strong dynamics~\cite{Harigaya:2016rwr,Co:2016akw}.
In this subsection, we review the dynamical generation of the dark photon mass.

Let us continue to consider a model with $N_f =2$ which has $ SU(3)_{D} \times U(1)_{D} $ gauge symmetry as in the previous subsection.
The charge assignment of $U(1)_D$ is, on the other hand, changed to the one in Tab.\,\ref{tab:tab chiral two flavor}. 
For $ 0<a < 1 $, the $U(1)_D$ gauge symmetry is no more vector-like symmetry, and hence, the 
mass terms in Eq.\,\eqref{eq:Lmass} are 
now forbidden.
The $U(1)_{D}$ gauge symmetry is free from gauge anomalies.
The global $U(1)_{B-L}$ symmetry, on the other hand, has the global anomaly of $ U(1)_{B-L} \times U(1)_{D}^{2}$, although this does not affect the ADM scenario unless there is a dark helical magnetic field in the Universe (see, e.g., Ref.~\cite{Kamada:2016cnb}).

The assumption of the chiral $U(1)_D$ is crucial for the dynamical breaking of $U(1)_D$,
since the vector-like symmetry cannot be broken spontaneously by strong dynamics~\cite{Vafa:1983tf}.
Note also that the $U(1)_D$ gauge symmetry explicitly breaks the $SU(2)'_{L} \times SU(2)'_{R} $ flavor symmetry of the dark quarks into the third component of the dark isospin symmetry, $U(1)_{3}'$.
Hereafter, $ I^{\prime}_{3} $ refers to the charge under $U(1)_{3}'$.
The $SU(2)'_{L} \times SU(2)'_{R} $ flavor symmetry remains an approximate symmetry as long as the $U(1)_D$ gauge interaction is perturbative.

\begin{table}[tb]
 \begin{center} 
  \begin{tabular}{|c|c|c|c|c|} \hline
           & $SU(3)_{D}$ & $U(1)_{D}$ & $U(1)_{B-L}$ & $U(1)_{3}'$ \\ \hline
    $U^{'}$ & $\mathbf{3}$ & 1 & 1/3 & 1 \\
    $D^{'}$ & $\mathbf{3}$ & $ -1 $ & 1/3 & $ -1 $ \\
    $\bar{U}^{'}$ & $\bar{\mathbf{3}}$ & $ -a $ & $ -1/3 $ & $ -1 $ \\ 
    $\bar{D}^{'}$ & $\bar{\mathbf{3}}$ & $ a $ & $ -1/3 $ & 1 \\ \hline
  \end{tabular}
   \caption{\small\sl The charge assignment of the chiral composite ADM model. We take $0< a < 1$, and hence, the $U(1)_D$ gauge symmetry is not vector-like.
   The $U(1)_{3}'$ symmetry is the third component of the dark isospin symmetry, $SU(2)'_V$.}
  \label{tab:tab chiral two flavor}
 \end{center} 
\end{table}

Below $ \Lambda_{D}'$, the dark quark bilinears 
condense as follows,
\begin{align}
\Braket{ U^{\prime} \bar{U}^{\prime} + U^{ \prime \dagger} \bar{U}^{ \prime \dagger} } = \Braket{ D^{\prime} \bar{D}^{\prime} + D^{ \prime \dagger} \bar{D}^{ \prime \dagger} } = \order{\Lambda_D^{\prime 3}}\ .
\end{align}
The condensate in this channel is expected to be favored than other channels such as $\langle U'\bar{D}'\rangle$
since this channel has the smallest $U(1)_D$ charge,
that is $|1-a|<|1+a|$, for $0<a<1$~\cite{Harigaya:2016rwr}.
These condensations spontaneously break the $U(1)_D$ gauge symmetry.
Besides, they also break the 
approximate $ SU(2)'_{L} \times SU(2)'_{R}$ flavor symmetry 
into the diagonal subgroup, $SU(2)'_{V}$.
On the other hand, $ U(1)_{B-L}$ is not broken by the condensations, and hence, $U(1)_{B-L}$ and $U(1)_{3}'$ are exact (accidental) symmetries up to $U(1)_D$ anomaly.%
\footnote{The appearance of the accidental symmetries is generic advantage of the composite dark matter models~\cite{Antipin:2015xia,Bottaro:2021aal}.}

Associated with spontaneous breaking of 
$SU(2)'_{L} \times SU(2)'_{R}$ into $SU(2)'_V$,
there are three pseudo Nambu-Goldstone (NG) bosons.
The low energy effective theory of the NG bosons is well described by the matrix-valued $SU(2)$ field, 
\begin{align}
\label{eq:U}
U(x) = \mathrm{exp} \Biggl[ \frac{i}{f_{\pi}'} \sum_{i=1}^{3} \pi'_{i}(x) \sigma_{i} \Biggr]\ ,
\end{align}
where $f_{\pi}'$ denotes the dark pion decay constant, $ \pi'_{i}(x) \, (i = 1,2,3) $ are three Nambu-Goldstone (NG)  bosons and $\sigma_i\, (i = 1,2,3)$ are the Pauli matrices.%
\footnote{We take the normalization of Eqs.\,\eqref{eq:U} and \eqref{eq:Leff U} so that the corresponding $f_\pi$ in the QCD is $f_\pi \simeq 93$\,MeV.}
Of these three NG bosons, $ \pi'_{3} $ becomes the longitudinal component of the dark photon, which is evident that $U(1)_D$ is realized 
by the shift of $\pi'_3$ at around $\Vec{\pi'}=0$.
Hereafter, we call the two remaining NG bosons, $\pi' \equiv (\pi'_{1} + i\pi'_{2} )/ \sqrt{2}$ (and its complex conjugate 
${\pi}^{\prime\dagger}$), 
the dark pions.

The kinetic term and the $U(1)_D$ gauge interaction of the dark pion is described by 
\begin{align}
\label{eq:Leff U}
\mathcal{L} = \frac{f^{\prime 2}_{\pi}}{4} \mathrm{tr} [ (D_{\mu}U) (D^{\mu}U)^{\dagger} ]\ ,
\end{align}
where the covariant derivative of $U(x)$ is given by,
\begin{align}
  D_{\mu}U = \partial_{\mu}U(x) 
    - i e_{D}A_{\mu}' \sigma_3 U(x)
    + i ae_{D}A_{\mu}'U(x) \sigma_3\ .
\end{align}
Here, $e_{D}$ is the gauge coupling constant of $ U(1)_{D} $.
In the ``unitary gauge",  $\pi'_3 = 0$,  we obtain interactions between the dark pion and the dark photon,
\begin{align}
\label{eq:Leff phi}
\mathcal{L} = (D_{\mu} \pi')^\dagger (D^{\mu} \pi') - e^{2}_{D} (1-a)^{2} A_{\mu}'A^{\prime \mu}\pi^{\prime\dagger}\pi' + 
\frac{1}{2} e_D^2(1-a)^2 f_\pi^{\prime 2} A_\mu^{\prime 2}
+\cdots \ ,
\end{align}
where the ellipses denote the 
higher dimensional terms suppressed by $f_{\pi'}$.
We introduced the ``covariant" derivative of $\phi$,  
\begin{align}
D_{\mu} \pi' = \partial_{\mu} \pi' + ie_{D} (1+a) A_{\mu}' \pi' \ .
\end{align}
The $U(1)_D$ invariance of Eq.\,\eqref{eq:Leff phi} 
is not manifest due to the non-linear realization of $U(1)_D$, although the effective theory in Eq.\,\eqref{eq:Leff U} is
manifestly $U(1)_D$ invariant. 

The third term of Eq.\,\eqref{eq:Leff phi} gives the dark photon mass,
\begin{align}
\label{eq:mass of dark photon}
m_{\gamma'} = e_{D} (1-a) f_{\pi}' \simeq \frac{\sqrt{3}}{4 \pi} e_{D} (1-a) m_{\rho'}\ ,
\end{align}
In the final expression, we have used the naive dimensional analysis between the (dark) pion decay constant and the (dark) rho meson mass~\cite{Manohar:1983md},
\begin{align}
    f_\pi' \simeq \frac{\sqrt{N_c}}{4\pi} m_{\rho'}\ ,
\end{align}
with $N_c = 3$.
Here, $ m_{\rho^\prime} $ is the mass of the dark rho meson.
The dark photon becomes massless for $a = 1$, which corresponds to the vector-like $U(1)_D$.

As the $U(1)_D$ gauge interaction breaks
the $SU(2)'_L\times SU(2)'_R$ symmetry explicitly, the corresponding NG boson $\pi'$ obtains the non-vanishing mass. Following \cite{Das:1967it} (see also \cite{Cheng:1985bj}),
we obtain the dark pion mass squared,%
\footnote{This expression is four times larger than that in Ref.~\cite{Harigaya:2016rwr}.}
\begin{align}
\label{eq:mass of dark pion}
m_{\pi^\prime}^{2} \simeq \frac{3a \log{2} }{2 \pi^{2}} e_{D}^{2} m_{\rho'}^{2}\, .
\end{align}
Here, we neglected the dark photon mass 
whose effects are suppressed by $\order{m_{\gamma'}^2/m_{\rho'}^2}$.
The dark pion becomes massless for $a=0$, where 
the $U(1)_D$ gauge symmetry does not break $SU(2)'_R$ explicitly.
By comparing Eqs.\,\eqref{eq:mass of dark photon}
and \eqref{eq:mass of dark pion}, we find that the condition $a\gtrsim 0.2$ is required for the dark pion mass to be larger than the dark photon mass.

\begin{figure}[tbp]
\centering{\includegraphics[width=0.6\textwidth]{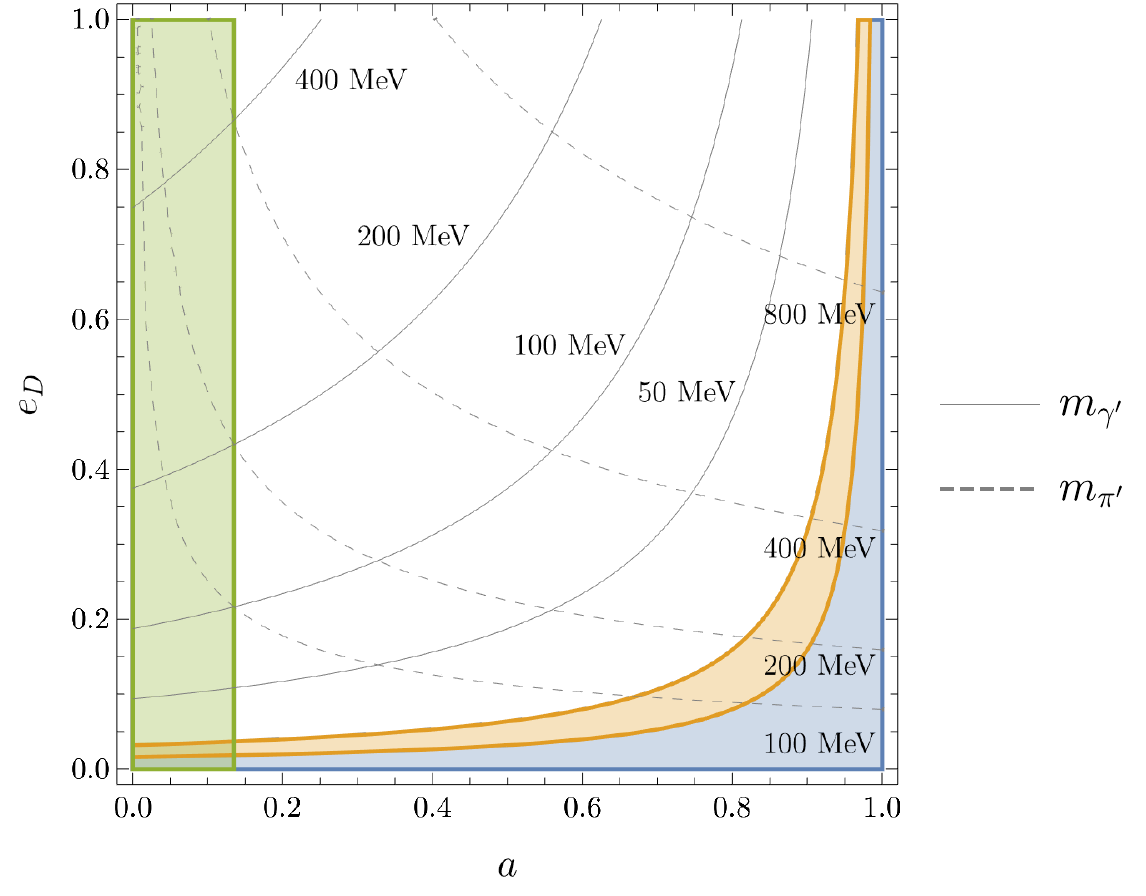}} 
\caption{
{
The contours of the dark photon/pion masses on a $(a,e_D)$ plane.
Here, we take $m_{\rho'} = m_{\rho}\times 5$ ($m_{\rho}\simeq 775$\,MeV), which corresponds to the dark baryon mass, $m_{p',n'}\simeq 5$--$6$\,GeV.
The green shaded region corresponding to $m_{\pi'} < m_{\gamma'}$ is disfavored.
The blue and the orange shaded regions
correspond to the $N_\mathrm{eff}$ constraints from the current
and the future CMB observations
for $\epsilon \gtrsim 10^{-9}$~\cite{Ibe:2019gpv}.
}
}
\label{fig:mass}
\end{figure}

{Some constraints on the dark photon/pion masses put bounds on $a$ and $e_D$.
In Fig.~\ref{fig:mass}, we show the viable parameter region.
The figure shows that the requirement $m_{\pi'} > m_{\gamma'}$ is achieved for $a\gtrsim 0.13$ (outside the green shaded region).
 We also show the lower limits on $m_{\gamma'}$ from the effective number of neutrino degrees of freedom, $N_{\mathrm{eff}}$~\cite{Ibe:2019gpv} as blue/orange shaded regions.
 The blue shaded region corresponding to
 $m_\mathrm{\gamma'} \lesssim 8.5$\,MeV is excluded by the $N_\mathrm{eff}$ constraint from the Planck observation of the cosmic microwave background (CMB)~\cite{Aghanim:2018eyx} for $\epsilon\gtrsim 10^{-9}$. 
 The orange one corresponding to $m_\mathrm{\gamma'} \lesssim 17$\,MeV shows the future sensitivity of the stage-IV CMB experiment~\cite{CMB-S4:2016ple}.
}

\renewcommand{\thesubsection}{\Alph{subsection}}
\subsection{Chiral Composite ADM for \texorpdfstring{$ N_{f} = 2 $}{} }
By combining the ideas in Sec.\,\ref{sec:CADM} and Sec.\,\ref{sec:Dynamical}, we construct 
the chiral composite ADM model in which the
dark photon mass is generated dynamically.\
In the model with $N_f =2$ so far considered,
the lightest dark baryons are
\begin{align}
p^{\prime} \propto U^{\prime} U^{\prime} D^{\prime} ~ , ~~ \bar{p}^{\prime} \propto \bar{U}^{\prime} \bar{U}^{\prime} \bar{D}^{\prime} ~ , ~~ 
n^{\prime} \propto U^{\prime} D^{\prime} D^{\prime} ~ , ~~ 
\bar{n}^{\prime} \propto \bar{U}^{\prime} \bar{D}^{\prime} \bar{D}^{\prime} ~ ,
\end{align}
as in QCD.
Due to the non-trivial baryon charges, they are stable and eventually become DM.
The mass partner of $p'$ is $\bar{p}'$, and that of $n'$  is $\bar{n}'$.
This combination results from the $U(1)'_3$ and the $U(1)_{B-L}$ symmetries.
Besides, the masses of the dark proton and
the dark neutron are identical due to the charge conjugation symmetry 
which interchanges $U'$ and $D'$.
Therefore, both the dark proton and the dark neutron become dark matter.

In summary, we obtain the chiral composite ADM model based on the charge assignment in Tab.\,\ref{tab:tab chiral two flavor}, which  
naturally provides the desirable spectrum,
\begin{align}
    m_{\gamma'} < m_{\pi'} < m_\mathrm{DM}\ .
\end{align}
In this model,
\begin{itemize}
    \item The dark baryons are stable due to $B-L$ symmetry.
    \item The dark baryon density is dominated by the asymmetric components due to the large annihilation cross section of the dark baryons into the dark pions.
    \item The chiral $U(1)_D$ gauge symmetry is broken dynamically, whose mass is suppressed compared with the dark baryons.
    \item The dark pion mass is generated by the radiative correction of the $U(1)_D$ interaction. The mass can be larger than the dark photon mass, and hence, the dark pion annihilates into the dark photon pairs efficiently.%
    \footnote{In the present model, there is no ``dark neutral pion" as it is absorbed by the dark photon. Thus, there is no need to assume that the dark pion mass is twice larger than the dark photon mass. }
\end{itemize}
For $A_{\mathrm{DM}}/A_{\mathrm{SM}}= \order{1}$, the dark matter density is reproduced by $m_\mathrm{DM} = \order{1}$\,GeV, 
with which the masses of the dark photon and the dark pion are predicted in the sub-GeV due to the dynamical symmetry breaking.

\renewcommand{\thesubsection}{\Alph{subsection}}
\subsection{Chiral Composite ADM for \texorpdfstring{$ N_{f} = 3 $}{} }
For successful ADM models, 
we need a portal interaction with which the 
$B-L$ asymmetry is thermally distributed between
the visible and the dark sectors.
One of such portal interactions can be given by
higher-dimensional operators~\cite{Ibe:2011hq,Fukuda:2014xqa},
\begin{align}
\label{eq:B-L portal O}
    \mathcal{L}_{B-L\,\mathrm{portal}} =  \frac{1}{M_*^{n}} \mathcal{O}_{D}LH\ + \mathrm{h.c.}
\end{align}
Here, $L$ and $H$ denote the lepton and the Higgs doublets in the visible sector. 
$\mathcal{O}_D$ is a $B-L$ charged but $U(1)_D$ neutral operator in the dark sector with the mass dimension $d_D = n + 3/2$.
$M_*$ encapsulates the energy scale of the portal interactions.
With this portal interaction, the visible and the dark sectors are in the thermal equilibrium at the high temperature.
The $B-L$ asymmetry is also thermally distributed through the same operators.
The portal interaction eventually decouples at the temperature,
\begin{align}
\label{eq:TD}
    T_D \sim M_* \left(\frac{M_*}{M_\mathrm{Pl}}\right)^{1/(2n-1)}\ ,
\end{align}
where $M_{\mathrm{Pl}}\simeq 2.4 \times 10^{18}$\,GeV is the reduced Planck mass scale.
Below the decoupling temperature, the $B-L$ asymmetries in the two sector are conserved separately.

In the model in the previous subsection, however, there is no good candidate for $\mathcal{O}_D$ which is $B-L$ charged
but neutral under the $SU(3)_D\times U(1)_D$ gauge symmetry.
To allow the $B-L$ portal interaction in Eq.\,\eqref{eq:B-L portal O}, we extend the composite ADM model with an additional generation of dark quarks so that $N_f = 3$.%
\footnote{We may instead extend the non-Abelian gauge group to $SU(2)_D$ while keeping $N_f = 2$.}
The charge assignment of the dark quarks under the gauge and the global symmetries are given in Tab.\,\ref{tab:Nf3}.
\begin{table}[t]
 \begin{center}
  \begin{tabular}{|c|c|c|c|c|c|} \hline
           & $SU(3)_{D}$ & $U(1)_{D}$ & $U(1)_{B-L}$ & $U(1)_{3}'$ & $U(1)_{8}'$ \\ \hline
    $U^{\prime}$ & \bf{3} & 1 & 1/3 & 1 & 1 \\
    $D^{\prime}$ & \bf{3} & $ -1 $ & 1/3 & $ -1 $ & 1 \\
    $S^{\prime}$ & \bf{3} & 0 & 1/3 & 0 & $ -2 $ \\
    $\bar{U}^{\prime}$ & $\bar{\bf{3}}$ & $ -a $ & $ -1/3 $ & $ -1 $ & $ -1 $ \\ 
    $\bar{D}^{\prime}$ & $\bar{\bf{3}}$ & $ a $ & $ -1/3 $ & 1 & $ -1 $ \\
    $\bar{S}^{\prime}$ & $\bar{\bf{3}}$ & 0 & $ -1/3 $ & 0 & 2 \\
    \hline
  \end{tabular}
  \caption{\small\sl The charge assignment of the dark quarks for $N_f = 3$.
  The $U(1)_{3}'$ and $U(1)_{8}'$ symmetries are the two components of the Cartan subgroup of the vector-like $SU(3)'_V$ symmetry.}
  \label{tab:Nf3}
 \end{center} 
\end{table}
In this model, the third flavor, $(S',\bar{S}')$, 
is vector-like, and hence, it has a mass $m_{S'}$,
where we assume $m_{S'} < \Lambda_D'$.
In the presence of the third flavor dark quark,
the $B-L$ portal interaction can be given by,
\begin{align}
\label{eq:B-L portal}
\mathcal{L}_{B-L\,\,\mathrm{portal}} = \frac{1}{M_{*}^{3}} ( {U}^{\prime} {D}^{\prime} {S}^{\prime} ) LH + \frac{1}{M_{*}^{3}} ( \bar{U}^{\prime \dagger} \bar{D}^{\prime \dagger} S^{\prime} ) LH + \mathrm{h.c.} 
\end{align}

{For the $B-L$ asymmetry generation mechanism, we assume, for example,
the thermal leptogenesis~\cite{Fukugita:1986hr} as in Ref.~\cite{Ibe:2018juk}.
In the thermal leptogenesis, the reheating temperature, $T_R$, after inflation is required to be higher than, $T_R \gtrsim 10^{9.5}$\,GeV~\cite{Giudice:2003jh,Buchmuller:2005eh}. Thus, for the scenario to be successful, we require $T_D \lesssim T_R$, which reads, 
\begin{align}
    M_* \lesssim T_R
    \left(\frac{M_\mathrm{Pl}}{T_R}\right)^{1/2n} 
    \lesssim 
    10^{11}\,\mathrm{GeV}
    \times 
    \left(\frac{T_R}{10^{9.5}\,\mathrm{GeV}}
    \right)^{5/6}\ ,
\end{align}
where we have used $n=3$ in the last inequality.
In this case, we assume that the portal interaction is generated by a 
new sector at $M_* \simeq 10^{11}$\,GeV (see Ref.~\cite{Ibe:2018juk} for details).}

{
Our portal mechanism works with 
other types of baryogenesis in the visible sector.
Again, $T_D$ should be lower than the temperature
of the completion of the baryogenesis. 
In addition, when the baryogenesis does not generate the lepton asymmetry directly, $T_D$ is required to be lower than 
the temperature
$T_\mathrm{sph}\sim 10^{12}$\,GeV, below which the sphaleron process is in equilibrium (see e.g. Ref.~\cite{Moore:2000ara}).
Other leptogenesis in the visible sector which completes above $T_D$
also works. 
However, the numerical value of the dark sector asymmetry 
will be altered from the one used in the present paper if $T_\mathrm{sph} < T_D$.
}

For $m_{S}'\lesssim \Lambda_D'$, the dark quark bilinears condense as,
\begin{align}
\label{eq:condensation}
\Braket{ U^{\prime} \bar{U}^{\prime} + U^{ \prime \dagger} \bar{U}^{ \prime \dagger} } = \Braket{ D^{\prime} \bar{D}^{\prime} + D^{ \prime \dagger} \bar{D}^{ \prime \dagger} } = \Braket{ S^{\prime} \bar{S}^{\prime} + S^{ \prime \dagger} \bar{S}^{ \prime \dagger} } = \order{\Lambda_D^{\prime 3}} \ .
\end{align}
The condensates spontaneously break the $U(1)_D$ gauge symmetry and the approximate $SU(3)'_L\times SU(3)'_R$ symmetry to the diagonal subgroup $SU(3)'_V$.
Note that the two components of the Cartan subgroup of $SU(3)'_V$, $U(1)_3'$ and $U(1)_8'$, are 
exact (accidental) symmetry 
up to $U(1)_D$ anomaly
and remain unbroken by the condensates in Eq.\,\eqref{eq:condensation}.
As in the model with $N_f = 2$, the NG boson 
corresponds to $\pi'_3$ becomes the longitudinal component of the $U(1)_D$ gauge boson.

As in QCD,
the dark quarks are confined into hadrons,
and the lightest baryons and the NG bosons
form the octet representations of $SU(3)'_V$,
\begin{align}
  B_{\alpha}^{\prime}\! = \!\left(
    \begin{array}{ccc}
      \Sigma^{\prime 3}_{\alpha} + \Lambda^{\prime}_{\alpha} / \sqrt{3} & \sqrt{2} \Sigma^{\prime 1}_{\alpha} & \sqrt{2} p_{\alpha}^{\prime} \\
      \sqrt{2} \Sigma^{\prime 2}_{\alpha} & -\Sigma^{\prime 3}_{\alpha} + \Lambda^{\prime}_{\alpha} / \sqrt{3} & \sqrt{2} n_{\alpha}^{\prime} \\
      \sqrt{2} \Xi^{\prime 2}_{\alpha} & \sqrt{2} \Xi^{\prime 1}_{\alpha} & -2 \Lambda^{\prime}_{\alpha} / \sqrt{3}
    \end{array}
  \right) , \,\,\,
  M^{\prime}\! =\! \left(
    \begin{array}{ccc}
      \eta^{\prime} / \sqrt{3} & \sqrt{2} \pi^{\prime} & \sqrt{2} K^{\prime 1} \\
      \sqrt{2} \pi^{\prime\dagger} & \eta^{\prime} / \sqrt{3} & \sqrt{2} K^{\prime 2} \\
      \sqrt{2} K^{\prime 1\dagger} & \sqrt{2} K^{\prime 2\dagger } & -2\eta^{\prime} / \sqrt{3}
    \end{array}
  \right)\ ,
  \end{align}
where we have taken the unitary gauge, i.e.,  $\pi'_3=0$.
The index $\alpha$ denotes the Weyl spinor components.
The names of the dark baryons and the dark mesons are after the corresponding baryons and the mesons in the visible sector.%
\footnote{The dark $\eta'$ corresponds to $\eta$ in the visible sector.}
The $U(1)_D$ charges of them are 
not parallel with 
the $U(1)_\mathrm{QED}$ charges in the visible sector.
The dark baryons are also associated with their antiparticles, $\bar{B}_\alpha$.
The mesons $\pi'$ and $K^{\prime 1,2}$ are complex scalars,
while $\eta'$ is a pseudo scalar.

As in the model with $N_f = 2$, all the physical NG bosons become massive.
The $\pi'$ mass is from the $U(1)_D$ interaction as in Eq.\,\eqref{eq:mass of dark pion}.
The NG bosons which have the $S$ components 
obtain masses of $\order{\sqrt{m_{S'}\Lambda_D'}}$ due to 
the explicit mass term of $(S',\bar{S}')$.%
\footnote{Due to the charge conjugation symmetry which exchanges $U'$ and $D'$, the masses of $K^{\prime 1,2}$ are identical.}
In the following, we assume that they are slightly heavier than the dark pion, 
so that they annihilate into the dark pions very efficiently.
Note that all the NG bosons other than $\eta'$ are stable 
when no NG bosons are twice heavier than the other
NG bosons.%
\footnote{When $\eta'$ is twice heavier than
$\gamma'$, it decays into a pair of the dark photons.
For a lighter $\eta'$, it decays into $\gamma' + e^++e^-$ through the kinetic mixing.
}

The leading mass term of the baryons 
is from
\begin{align}
\label{eq:Baryon mass}
    \mathcal{L}_{\mathrm{mass}} &\simeq \frac{1}{2} m_B \tr[BU\bar{B}U^\dagger] + \mathrm{h.c.} \ ,\\
    &\simeq 
    m_B(p'\bar{p}'+n'\bar{n}'+ \Lambda'\bar{\Lambda}'
    + \sum_{i=1}^3\Sigma^{\prime i}\bar{\Sigma}^{\prime i} 
    + \sum_{i=1}^2\Xi^{\prime i}\bar{\Xi}^{\prime i} 
    ) + \mathrm{h.c.}\ ,
\end{align}
where
\begin{align}
     U(x)  = \exp\left[\frac{i}{f_\pi'} M(x)\right]\ .
\end{align}
The term in Eq.\,\eqref{eq:Baryon mass}
is invariant under $U(1)_D\times SU(3)'_L\times SU(3)'_R$.
Note that the dark baryon masses are not identical due to the $U(1)_D$ gauge interaction and the mass of $(S',\bar{S}')$.

\section{Phenomenology and Cosmology of \texorpdfstring{$N_f = 3$}{} Model}
\label{sec:phenomenology}
When the mass difference between the dark baryons larger than the dark NG boson masses, the heavier 
baryons decay into the lighter baryons by emitting a dark NG boson.
Besides, $ \Sigma^{\prime 3} $ decays into a pair of
$\Lambda^{\prime}$ and the dark photon when their mass difference is larger than the mass of the dark photon.
Even if the mass difference is smaller than the dark photon mass,  $ \Sigma^{\prime 3} $ eventually decays
into $\Lambda^{\prime}+e^++e^-$ through the kinetic mixing of the dark photon.
As the total dark baryon number is conserved, however,
the details of the decay properties of the dark baryons do not affect the number density of the dark baryons.
Since all the dark baryon masses are around $m_B$, the dark baryon mass density is also insensitive to the details of the decay properties.

Here, let us comment on the dark baryon density.
As shown in Ref.~\cite{Fukuda:2014xqa},
the ratio $A_{\mathrm{DM}}/A_\mathrm{SM}$ is given by
\begin{align}
    \frac{A_\mathrm{DM}}{A_\mathrm{SM}} = \frac{22}{237}N_f\ ,
\end{align}
when $T_D$ in Eq.\,\eqref{eq:TD} is lower than the temperature $T_e$
at which the electron Yukawa coupling becomes effective
in the thermal bath.
When $T_D$ is higher than $T_e$ but lower than the temperature $T_{ud}$ at which the up and down Yukawa couplings become effective, the ratio is slightly changed to 
\begin{align}
    \frac{A_\mathrm{DM}}{A_\mathrm{SM}} = \frac{20}{213}N_f\ ,
\end{align}
(see the Appendix \ref{sec:calc of asym}).
When $T_D$ is higher than $T_{ud}$, the ratio is given by,
\begin{align}
    \frac{A_\mathrm{DM}}{A_\mathrm{SM}} = \frac{17}{149}N_f \ .
\end{align}
From these ratios and Eq.\,\eqref{eq:DMmass}, we find that the DM mass is predicted to be $m_\mathrm{DM}\simeq 5$--$6$\,GeV for $N_f = 3$
for wide range of $T_D$.
In order to achieve this $m_\mathrm{DM}$, we take $\Lambda_{D}'=\order{1}$\,GeV.

Note that each dark baryon decays into a pair of a dark meson and an anti-neutrino in the visible sector through the portal interaction of Eq.\eqref{eq:B-L portal}.
Thus, the dark baryons are not absolutely stable,
although their lifetime is longer than the age of the Universe for $M_* \gtrsim 10^{7.7}\,$GeV$\times(m_\mathrm{DM}/5\,\mathrm{GeV})^{1/2}$~\cite{Fukuda:2014xqa}.
From the the upper limit on the anti-neutrino flux over the predicted atmospheric flux measured by the Super-Kamiokande experiment~\cite{Desai:2004pq}, we require $M_* \gtrsim 10^{8.2}$\,GeV~\cite{Covi:2009xn,Fukuda:2014xqa}
for $m_\mathrm{DM} = 5$--$6$\,GeV,
which corresponds to the lifetime, 
$\tau_\mathrm{DM}\gtrsim 10^{21}$\,sec.

When the mass difference between the dark baryons are smaller than the dark NG boson masses,
multiple dark baryons with different masses and different $U(1)_D$ charges
become dark matter. 
The dark baryons with non-vanishing $U(1)_D$ charges can be tested by the direct detection signal by  exchanging the dark photon with the proton in the visible sector.
When the dark $U(1)_D$ gauge coupling constant is equal to QED coupling constant, a large portion of the parameter space can be tested by the future 
 XENONnT~\cite{Aprile:2020vtw}, LZ~\cite{Mount:2017qzi} and DARWIN~\cite{Aalbers:2016jon}
 experiments for $m_{\gamma'} \lesssim 100$\,MeV~\cite{Ibe:2018juk}.

Finally, we discuss the constraints on the dark NG boson
density.
As we mentioned earlier, most of the dark NG bosons 
are stable.
When the NG bosons containing $S'$ are slightly heavier than the dark pions, the heavier NG bosons annihilate
into the dark pions, while the dark pions annihilate into a pair of dark photons.%
\footnote{The dark photon decays before the neutrino decoupling temperature (see Ref.~\cite{Ibe:2019gpv}).}
Accordingly, the dark pion has the largest number density, and hence, we concentrate on the constraint on the dark pion density.

The thermally averaged annihilation cross-section of the dark pion is given by
\begin{align}
\label{eq:cross section}
\langle \sigma v \rangle = \frac{\pi \alpha^{2}_{D}}{m_{\pi'}^{2}} \mathcal{F}\biggl( \frac{m_{\gamma'}}{m_{\pi'}} \biggr) \ ,
\end{align}
where $\alpha_D = e_D^2/4\pi$ and
\begin{align}
 \mathcal{F}(x) = 16\frac{\sqrt{1-x^{2}}}{x^{4} (2-x^{2})^{2}}
 [&(a^{4}+1)(x^{2}-1)^{2}+ 2a(a^{2}+1)(x^{6}-3x^{4}+4x^{2}-2) \notag\\
&+3a^{2}(x^{8}-4x^{6}+6x^{4}-4x^{2}+2)]\ .
\end{align}
At first glance, this formula looks divergent if $m_{\gamma'}=0$.
However, this limit corresponds to $a\to 1$, then the formula becomes finite and reproduces the massless $U(1)$ gauge theory.
With this annihilation cross-section, the mass density of the dark pion is given by,
\begin{align}
    \Omega_{\pi'} \sim \frac{3\times 10^{-26}\,\mathrm{cm}^3 \,\mathrm{sec}^{-1}}{\langle \sigma v \rangle}\times \Omega_\mathrm{DM}\ ,
\end{align}
where we have used the WIMP cross-section, $\sigma v \sim 3\times 10^{-26}$\,cm$^3$\,sec$^{-1}$, with which the observed dark matter abundance is achieved by the freeze-out mechanism.%
\footnote{Here, we assume that there is no asymmetry between the dark pion and the dark anti-pion, which is justified in the Appendix\,\ref{sec:calc of asym}.}

If the relic abundance of the dark pion is sizable, 
the late time annihilation 
injects extra energy into the galactic medium 
after the recombination time.
The anisotropies of the cosmic microwave background (CMB) are sensitive to such energy injection, which put severe constraint on the 
energy injection rate~\cite{Aghanim:2018eyx}. 
The effective parameter constrained by the CMB anisotropies is given by
\begin{align}
\label{eq:pann}
p_{\mathrm{ann}} 
 &= f_{\mathrm{eff}} \frac{\langle \sigma v \rangle}{m_{\pi'}}\times \biggl(\frac{\Omega_{\pi'}}{\Omega_{\mathrm{DM}}}\biggr)^{2} 
 \simeq f_{\mathrm{eff}} \frac{\langle \sigma v \rangle}{m_{\pi'}}\times \biggl(\frac{3\times 10^{-26}\,\mathrm{cm}^3 \,\mathrm{sec}^{-1}}{\langle \sigma v \rangle}\biggr)^{2} 
 \ .
\end{align}
Here, the scaling factor $(\Omega_{\pi'}/\Omega_\mathrm{DM})^2$ comes from the fact that the energy injection rate is proportional to $m_{\pi'} n_{\pi'}^2 = \rho_{\pi'}^2/m_{\pi'}$ where $n_{\pi'}$ and $\rho_{\pi'}$ are the number and the energy densities of the dark pion, respectively.
$ f_{\mathrm{eff}} $ is an energy fraction released 
into the intergalactic medium around the red-shift $z\simeq 600$~\cite{Finkbeiner_2012}.
The dark photons produced by the dark pion annihilation eventually decays into a pair of the electron and the positron, and hence, $f_\mathrm{eff}  = \order{1}$~\cite{Slatyer:2015jla}.
By substituting \eqref{eq:cross section}, we obtain 
\begin{align}
p_{\mathrm{ann}} \simeq \frac{(m_{\pi'}/\mathrm{GeV})}{\pi \alpha^{2}_{D} \mathcal{F}(m_{\gamma'}/m_{\pi'})} \times 10^{-34} \, \mathrm{cm^{3}\,s^{-1}\,GeV^{-1}}\ .
\end{align}
For $m_{\pi'} > m_{\gamma'}$, $ \mathcal{F}(m_{\gamma'}/m_{\pi'}) $ is $\order{1}$ or larger, and hence, the 
dark pion density satisfies the current CMB constraint~\cite{Aghanim:2018eyx},
\begin{align}
    p_{\mathrm{ann}} < 3.5 \times 10^{-28}\,\mathrm{cm^{3}\,s^{-1}\,GeV^{-1}}\ .
\end{align}
Therefore, we find that the stable NG bosons do not cause observational problems.

The strong coupling between the dark nucleons and the dark pion does not lead to the self-interacting dark matter due to the velocity suppressed cross section~\cite{Chu:2018fzy}.
The dark photon exchange, on the other hand, causes the velocity independent self-interaction cross section $\sigma_0/m_{\mathrm{DM}} \sim 4\pi\alpha_D^2m_{\mathrm{DM}}/m_{\gamma'}^4 $, which is of $\order{0.01\mbox{--}0.1}$\,cm$^2$/g. 
This cross section is consistent with the constraints, $\sigma_0/m_{\mathrm{DM}}
\lesssim 0.1\mbox{--}1$cm$^2$/g, obtained from the galaxy clusters~\cite{Markevitch_2004,Randall_2008,Kahlhoefer_2013,Harvey_2015,Robertson_2016,Wittman_2018,Harvey_2019,Bondarenko:2020mpf,Sagunski:2020spe},
and, $\sigma_0/m_{\mathrm{DM}}
\lesssim 0.01\mbox{--}0.1$cm$^2$/g, obtained from the ultra-faint dwarf galaxies~\cite{Hayashi:2020syu,Zoutendijk:2021kee}.%
\footnote{There are also arguments in favor of finite interaction cross sections from the the dwarf irregular galaxies~\cite{Oh_2011}, the low surface brightness galaxies~\cite{de_Naray_2008}, and galaxy clusters~\cite{Newman_2013} (see also Ref.~\cite{Kaplinghat_2016,Kamada_2017,Kamada:2020buc}).}

\renewcommand{\thesection}{\Roman{section}}
\section{Conclusions}
\label{sec:conclusions}

Composite ADM models based on QCD-like strong dynamics are particularly motivated since the strong dynamics naturally provides the DM mass of $\order{1}$\,GeV and the large annihilation cross-section simultaneously. 
In this paper, we constructed 
a chiral composite ADM model where the $U(1)_D$ gauge symmetry is embedded into the chiral flavor symmetry. 
Due to the dynamical breaking of the chiral flavor symmetry, the model naturally provides the masses of the dark photon and the dark pions in the sub-GeV range, both of which play crucial roles for successful ADM models.
Let us emphasize that the dark photon mass is determined by the dynamical scale, which is 
an attractive feature of the present model compared with 
models with an additional Higgs boson to break $U(1)_D$ spontaneously.

The model with $N_f =3$ fits well with the scenario
where the $B-L$ asymmetry in the visible sector 
is thermally distributed to the dark sector through 
higher-dimensional $B-L$ portal operators~\cite{Ibe:2011hq,Ibe:2018juk}.
This type of scenario can be tested by multiple channels such as the direct detection~\cite{Ibe:2018juk}, the anti-neutrino flux from the decay of the dark baryons~\cite{Fukuda:2014xqa}, and the electron/positron flux from the annihilation of the dark baryons and the dark anti-baryons through the late-time oscillation~\cite{Ibe:2019yra}.

Let us also comment on the possibility 
of the first-order phase transition of the dark QCD.
In the present model, the chiral $U(1)_D$ gauge symmetry forbids
the dark quark masses of $U'$ and $D'$. 
The QCD with the vanishing up and down quark masses can exhibit the first-order phase transition depending on the strange quark mass~(see, e.g., \cite{Laermann:2003cv,deForcrand:2006pv}).
Thus, the dark QCD in the present model
may also have the first-order phase transition,
although the dark pion mass induced by the $U(1)_D$ gauge interaction could also
affect the order of the phase transition. 
Once the dark QCD undergoes the first order 
transition at the GeV range, 
the gravitational waves generated at the transition could be observed by the pulsar timing array experiments~(see, e.g., \cite{Schwaller:2015tja,Nakai:2020oit})
as well as the gravitational wave detection experiments~\cite{Huang:2020mso}.

{Finally, note that the baryon-DM coincidence problem is not fully solved by the ADM scenario without specifying the origin of the dark matter mass.
In fact, the puzzle is divided into two subproblems, which are the coincidence of masses and that of the number densities between baryons and DM. 
The ADM scenario naturally explains the coincidence of the number density while it does not answers the mass coincidence.
The composite ADM ameliorate the mass coincidence problem as it provides the dark matter mass via the dynamical transmutation in the dark QCD. However, it does not 
answer the coincidence problem unless the gauge coupling of the dark QCD is related to that of QCD. 
In Ref.~\cite{Ibe:2019ena}, we introduced a mirror symmetry under which the dark sector and the visible sector are exchanged so that the gauge coupling constants in the two sectors are related with each other.
In the present model, however, introducing the exchanging symmetry is difficult because the $U(1)_D$ gauge symmetry is chiral.
One possibility is to embed the dark QCD and dark QED into a chiral non-Abelian gauge theory, although we have not succeeded in constructing a concrete example.
We leave this issue for a future work.
}

\section*{Acknowledgements}
The authors thank S.~Shirai and K.~Harigaya for useful comments and correspondence.
This work is supported by Grant-in-Aid for Scientific Research from the Ministry of Education, Culture, Sports, Science, and Technology (MEXT), Japan, 17H02878, 18H05542 (M.I.), and by World Premier International Research Center Initiative (WPI), MEXT, Japan. 
This work is also supported by the Advanced Leading Graduate Course for Photon Science and the JSPS Research Fellowships for Young Scientists (S.K.).

\appendix
\section{Dark Pion Mass}
Since the $U(1)_D$ gauge symmetry forbids the mass term of the $U'$ and $D'$ dark quarks, the mass of the dark pion is generated by the $U(1)_D$ gauge interaction which breaks the $SU(2)'_L\times SU(2)'_R$ symmetry explicitly.
At the leading order of the $U(1)_D$ gauge coupling, the masses of the dark pions are given by~\cite{Cheng:1985bj,Das:1967it},
\begin{align}
    m_{a}^2 \delta^{ab} = \frac{e_D^2}{2f_{\pi'}^2}
    \int d^4x D_{\mu\nu}(x) \langle
[Q_A^{a}[Q_A^{b},T(j_D^\mu(x)j_D^{\nu}(0))]]
    \rangle\ .
\end{align}
Here, $D_{\mu\nu}$ is the dark photon propagator, and $Q_A^a (a=1,2,3)$ is the axial charges of $SU(2)'_L\times SU(2)'_R$ symmetry.
Since the dark pion is defined by $\pi' = (\pi'_1+i\pi'_2)/\sqrt{2}$, we take $a=b=1$ in the following.
The decay constant $f_{\pi'}$ is defined so that it corresponds to $f_\pi\simeq 93$\,MeV in the visible sector.
The $U(1)_D$ current is given by,
\begin{align}
    &j_D^\mu = \alpha j_V^{3\mu} + \beta j_A^{3\mu}\ , \\
&j_V^{3\mu} = \frac{1}{2}U'^\dagger \bar{\sigma}^\mu U' - \frac{1}{2}D'^\dagger \bar{\sigma}^\mu D' 
- \frac{1}{2}\bar{U'}^\dagger \bar{\sigma}^\mu \bar{U'} + \frac{1}{2}\bar{D'}^\dagger \bar{\sigma}^\mu \bar{D'}\ , \\
&j_A^{3\mu} = \frac{1}{2}U'^\dagger \bar{\sigma}^\mu U' - \frac{1}{2}D'^\dagger \bar{\sigma}^\mu D' 
+ \frac{1}{2}\bar{U'}^\dagger \bar{\sigma}^\mu \bar{U'} - \frac{1}{2}\bar{D'}^\dagger \bar{\sigma}^\mu \bar{D'}\ , \\
&\alpha = 1+a\ , \quad \beta = 1- a \ .
\end{align}
By using the commutation relations between $Q_A^1$ and $j_{A,V}^{3\mu}$, we obtain,
\begin{align}
    &m_{\pi'}^2 = \frac{4ae_D^2}{f_{\pi'}^2}
    \int d^4x D_{\mu\nu}(x)(V^{\mu\nu}(x)-A^{\mu\nu}(x))\ , \\
    &V^{\mu\nu}(x) = \langle T j_V^{3\mu}(x) j_V^{3\nu}(0)\rangle\ , \\
    &A^{\mu\nu}(x) = \langle T j_A^{3\mu}(x) j_A^{3\nu}(0)\rangle\ .
\end{align}
Then, following Ref.~\cite{Das:1967it}, we obtain,
\begin{align}
    m_{\pi'}^2 \simeq \frac{3a\log 2}{2\pi^2} e_D^2 m_{\rho'}^2\ ,
\end{align}
where we have neglected the dark photon mass in the propagator.

\section{Calculation of Asymmetry }
\label{sec:calc of asym}
In this appendix, 
we calculate the ratio of the $B-L$ asymmetries 
in the dark and the visible sectors, $A_{\mathrm{DM}}/A_{\mathrm{SM}} $, following Ref.~\cite{Weinberg:2008zzc}.
We also calculate the asymmetry between the 
dark pion and the dark anti-pion.

Let $ q_{ia} $ be a charge of a massless particle species $i$ in thermal equilibrium, where $ a $ denotes a conserved quantum number such as $ B-L $ or the weak hypercharge, $Y$. 
The chemical potential of the particle $i$, $\mu_{i}$, can be written as 
\begin{align}
\mu_{i} = \sum_{a} q_{ia} \mu_{a},
\end{align}
where $ \mu_{a} $ is a chemical potential associated with the conserved quantity. 
The difference between the number density of $i$ and its antiparticle at temperature $ T $ is given by,
\begin{align}
n_i - \bar{n}_i = \frac{T^2}{6} \tilde{g}_i \mu_i\, . 
\end{align}
Here, $ \tilde{g_{i}} $ is (twice of) a spin degree of freedom for a fermion (boson).
From these equations, the following equation holds,
\begin{align}
n_i - \bar{n}_i = \sum_{a,b} \tilde{g}_i q_{ia} M_{ab}^{-1} A_b\, ,
\end{align}
where $M_{ab} = \sum_{i} \tilde{g}_i q_{ia} q_{ib} $ and $ A_{a} =\sum_{i}A_{ia}= \sum_{i} q_{ia} (n_{i} - \bar{n}_i) $. 
Thus, by giving the asymmetries of the conserved quantum number, $A_a$, we obtain the particle-antiparticle asymmetries of each particles.

\subsection*{Asymmetry Ratio For
\texorpdfstring{$ T_{ {e} } > T_{D}  $}{}}
When the decoupling temperature of the $B-L$ portal interaction in Eq.\,\eqref{eq:TD}, $T_D$, is lower than the 
temperature $T_e$ at which the electron Yukawa coupling becomes effective, the ratio of the $B-L$ asymmetries 
are given by~\cite{Fukuda:2014xqa},
\begin{align}
\frac{A_{\mathrm{DM}}}{A_{\mathrm{SM}}}=\frac{22}{237}N_f\ .
\end{align}
Note that $U(1)_D$ gauge symmetry is conserved at $T\gg\Lambda'_D$ in the present model while there is no $U(1)_D$ gauge symmetry in the model in Ref.~\cite{Fukuda:2014xqa}.
With the additional conserved quantity,
$A_{\mathrm{DM}}/A_{\mathrm{SM}}$
can be different in general.
In the present model, however, 
$M_{B-L,U(1)_D}^{-1} = 0$,
and hence, the ratio $A_{\mathrm{DM}}/A_{\mathrm{SM}}$
coincides with the model in Ref.~\cite{Fukuda:2014xqa}.
For the model in Sec.\,\ref{sec:CADM}, on the other hand, $M_{B-L,U(1)_D}^{-1} \neq 0$,
and hence, the asymmetry ratio is slightly changed to $A_{\mathrm{DM}}/A_{\mathrm{SM}}=66N_{f}/395 $.

\subsection*{Asymmetry Ratio For
\texorpdfstring{$ T_{ {ud} } > T_{D} > T_{ \mathrm{e} } $}{}}
When $T_D$ is higher than $T_e$, we have an additional conserved 
quantum number, i.e.,  the number of the right-handed electron number in the visible sector.
The presence of the additional conserved quantum number alters the $B-L$ ratio.
By assuming the initial condition, $\mu_{e_R} = 0$,
we obtain, 
\begin{align}
    \frac{A_{\mathrm{DM}}}{A_{\mathrm{SM}}} = \frac{20}{213} N_{f}\ .
\end{align}
After the decoupling of the $ B-L $ portal, this value does not change, since the $ B-L $ charge is conserved separately in the dark and the visible sectors.

\subsection*{Asymmetry Ratio For \texorpdfstring{$ T_{D} > T_{ud} $}{}}
When the up and down Yukawa couplings are ineffective,
we also have an additional conserved quantum number in the visible sector corresponding to the charge under the $ U(1)_{T_{3R}} $ symmetry, which is the third component of the $SU(2)_R$ in the visible sector.
In this case, the asymmetry ratio becomes  
\begin{align}
\frac{A_{\mathrm{DM}}}{A_{\mathrm{SM}}} = \frac{17}{149} N_{f}\ .
\end{align}
As in the case of $T_{ud}>T_D>T_e$, this value does not change after the decoupling of the $ B-L $ portal.

\subsection*{Asymmetry of the Dark Pions}
Since we assume that only the $B-L$ asymmetry is generated,
the asymmetry between the dark pion and dark anti-pion is given by,
\begin{align}
    n_{\pi'} - n_{\bar{\pi}'} = 2 q_{\pi',I_3'} M_{I_3'B-L}^{-1}A_{B-L}\ ,
\end{align}
since $q_{\pi',B-L}=0$.
In the chiral ADM in Tab.\,\ref{tab:tab chiral two flavor} and in Tab.\,\ref{tab:Nf3},
we find that $M_{I_3'B-L}^{-1} = 0$, and hence, the dark pion does not have asymmetry.

\bibliography{bibtex}

\end{document}